\pdfoutput=1

\documentclass[journal,a4paper,10pt]{IEEEtran}
\usepackage[latin1]{inputenc}
\usepackage{times,amsmath}
\usepackage{amsfonts}
\usepackage{pstool}
\usepackage{subfigure}
\usepackage{multirow}
\usepackage{enumerate}
\usepackage{graphicx}
\usepackage{MnSymbol}
\usepackage{stfloats}
\usepackage[table]{xcolor}
\usepackage[square, comma, sort&compress, numbers]{natbib}
\usepackage{nohyperref}
\usepackage{algorithm,algorithmic}
\usepackage{bigints}
\usepackage{natbib}
\newcounter{tempEquationCounter}
\newcounter{thisEquationNumber}

\makeatletter
\newcommand{\vast}{\bBigg@{4}}
\newcommand{\Vast}{\bBigg@{5}}
\makeatother

\graphicspath{ {Figures/} }

\begin{document}

\title{Modulation Mode Detection \& Classification for in-Vivo Nano-Scale Communication Systems Operating in Terahertz Band}

\author{ M. Ozair Iqbal\authorrefmark{1}, M. Mahboob Ur Rahman\authorrefmark{1}, M. Ali Imran\authorrefmark{2}, Akram Alomainy\authorrefmark{3}, Qammer H. Abbasi\authorrefmark{2} \\
\authorblockA{
\authorrefmark{1}Department of Electrical engineering, Information Technology University, Lahore, Pakistan} \\
\authorblockA{
\authorrefmark{2}Department of Electronics and Nano engineering, University of Glasgow, Glasgow, UK} \\
\authorblockA{
\authorrefmark{3}School of Electronic Engineering and Computer Science, Queen Mary University of London, UK} \\
}

\maketitle

\begin{abstract} 

This work initiates the efforts to design an intelligent/cognitive nano receiver operating in Terahertz (THz) band. Specifically, we investigate two essential ingredients of an intelligent nano receiver---modulation mode detection (to differentiate between pulse based modulation and carrier based modulation), and modulation classification (to identify the exact modulation scheme in use). To implement modulation mode detection, we construct a binary hypothesis test in nano-receiver's passband, and provide closed-form expressions for the two error probabilities. As for modulation classification, we aim to represent the received signal of interest by a Gaussian mixture model (GMM). This necessitates the explicit estimation of the THz channel impulse response, and its subsequent compensation (via deconvolution). We then learn the GMM parameters via Expectation-Maximization algorithm. We then do Gaussian approximation of each mixture density to compute symmetric Kullback-Leibler divergence in order to differentiate between various modulation schemes (i.e., $M$-ary phase shift keying, $M$-ary quadrature amplitude modulation). The simulation results on mode detection indicate that there exists a unique Pareto-optimal point (for both SNR and the decision threshold) where both error probabilities are minimized. The main takeaway message by the simulation results on modulation classification is that for a pre-specified probability of correct classification, higher SNR is required to correctly identify a higher order modulation scheme. 

On a broader note, this work should trigger the interest of the community in the design of intelligent/cognitive nano receivers (capable of performing various intelligent tasks, e.g., modulation prediction etc.).

\end{abstract}

\section{Introduction}
\label{sec:intro}

Terahertz (THz) band ($0.1-10$ THz) is well-known as a promising candidate for nano-scale communication \cite{Akyildiz:Elsevier:2010,Ian:Nanonetworks:2008}, because of its non-ionization and robustness to the fading characteristics. Recently, the advent of novel nano-materials (e.g., graphene-based carbon nano tubes etc.) with remarkable electrical properties has led to a surge of interest in development of nano-scale devices \cite{Akyildiz:Elsevier:2010,Ian:Nanonetworks:2008} for communication in THz band. The literature so far comprises of the studies which investigate the antenna design \cite{Mona:BodyNets:2015}, propagation models \cite{Josep:TWC:2011},\cite{Qammer:TTHz:2016}, transceiver design \cite{Gupta:Globecom:2015},\cite{Ian:ICC:2012},\cite{Josep:TC:2014}, networking issues \cite{Jornet:Elsevier:2012,Han:CN:2013}, and data rates \cite{Josep:TWC:2011}. Nevertheless, the open literature on nano-scale communication in THz band still falls short of the problem of intelligent nano receiver design. Therefore, this work initiates the efforts to design an intelligent/cognitive nano receiver operating in Terahertz (THz) band. 

Intelligent/cognitive receiver design for nano-scale communication in THz band finds its utilization in defense, security and military applications \cite{Iwaszczuk:PhDThesis:2012}. Modulation mode detection (where one discriminates between the pulse based modulation and the carrier based modulation), and modulation classification (where one identifies the modulation scheme in use) constitute two essential functions of an intelligent nano receiver, among others. Specifically, a cognitive nano receiver performs various statistical inference tests on the received signal of interest to implement modulation mode detection and modulation classification in a systematic manner.

The first essential ingredient of an intelligent nano receiver that this work investigates is modulation mode detection. The need for modulation mode detection arises due to the fact that the nano-scale communication systems operating in THz band either do (few hundred femto-seconds long) pulse-based communication, or, utilize classical carrier-based modulation schemes (if form factor is not a constraint) \cite{Josep:TC:2014},\cite{MOSHIR:NCN:2016}. This implies that the nano transmitter of interest could have dual modes of modulation; i.e., it could switch between pulse based modulation and carrier based modulation on slot basis. This in turn necessitates that an intelligent nano receiver should build a systematic mechanism to detect the mode of modulation upon reception of the signal of interest, during every slot. The nano receiver then utilizes the outcome of mode detection to activate one of the two decoding chains (i.e., pulse based decoding, carrier based decoding). 

The second essential ingredient of an intelligent nano receiver that this work considers is modulation classification. As the name implies, modulation classification automatically identifies the modulation scheme in use, from the received signal of interest. It is the intermediate step between signal detection and demodulation. Modulation classification has been extensively studied in the literature on traditional wireless networks operating on microwave frequencies (see the survey article \cite{Octavia:IETComm:2007} and references therein). There, it was originally motivated by military applications; but now, it finds its application in various cooperative communication problems, e.g., cognitive radios etc. 

For modulation classification in traditional wireless networks, the solutions reported in the literature so far could be broadly classified as either feature-based (e.g., cyclic cumulants, moments, amplitude, phase etc.), or, decision-theoretic (based on likelihood functions) \cite{Octavia:IETComm:2007},\cite{Wang:Globecom:2012}. Decision-theoretic approaches are optimal, but are computationally prohibitive, and sensitive to model mismatch (e.g., frequency, timing offsets etc.). Pattern recognition based approaches, on the other hand, could perform very close to optimal if designed properly. For feature-based approaches, cumulants are generally preferred over moments whereby up to 8-th order cumulants have been reported in the literature \cite{Wu:TWC:2008}. More recently, there is a growing interest in applying machine learning techniques to automatic modulation recognition/classification: e.g., convolutional neural networks based deep learning \cite{Peng:WOCC:2017}, deep neural networks \cite{Ali:PhyComm:2017}, $K$-nearest neighbors approach, and support vector machine \cite{Zhu:WileyCh6:2014}.

{\it Nevertheless, to the best of the authors' knowledge, modulation mode detection as well as modulation classification--- or, broadly speaking--- the design of intelligent/cognitive nano receiver have not been addressed so far in the open literature on nano-scale communication in THz band.} \\

On a related note, the design of efficient modulation schemes with various design objectives (e.g., energy efficiency, interference analysis in a multi-users scenario etc.) for nano-scale communication systems operating in THz band has attracted researchers' interest very recently. In \cite{MOSHIR:NCN:2016}, authors propose a pulse based modulation scheme, and a rate adaptation scheme to take into account the highly frequency-selective nature of THz band and signal attenuation due to molecular absorption. \cite{MOSHIR:NCN:2016} also provides a comprehensive summary of the various modulation schemes used in (simulation based and prototype) THz systems so far (see Table 1 in \cite{MOSHIR:NCN:2016}). Earlier, Josep et. al. in \cite{Josep:TC:2014} considered the design of time-spread on-off keying (TS-OOK) scheme---a pulse based modulation scheme---and evaluated the data rates in single user scenario and multiple users scenario (under time-division multiplexing strategy) respectively. In \cite{Vavouris:MOCAST:2018}, authors utilize TS-OOK scheme to realize energy-efficient body-centric nano-scale communication systems operating in THz band. Finally, Abbasi et. al. in \cite{Zhang:Access:2017} consider an in-vivo multi-user (and hence interference-limited) scenario where the nano transmitter employs the TS-OOK scheme; authors compute the distribution of the signal-to-interference-plus-noise ratio for communication through various human tissues (e.g., blood, skin, fat etc.) to compute the achievable communication distance.

{\bf Contributions.} This work considers the design of an intelligent/cognitive nano receiver operating in THz band. Specifically, this work has the following distinct contributions:

\begin{itemize}

\item We do modulation mode detection to systematically differentiate between pulse based modulation and carrier based modulation. To this end, we formulate the problem at hand as a binary hypothesis test in receiver passband, and provide closed-form expressions for the two error probabilities. 

\item When modulation mode detection declares carrier based modulation, we do feature-based modulation classification to identify the specific modulation scheme in use. Two kinds of modulation schemes---$M$-ary phase shift keying ($M$-PSK), and $M$-ary quadrature amplitude modulation ($M$-QAM)---are considered, which are fundamental, ubiquitous and practical modulation schemes. As for modulation classification, we represent the received signal of interest by a Gaussian mixture model, and learn its parameters via expectation maximization algorithm. We then do Gaussian approximation of each mixture density to invoke symmetric Kullback-Leibler divergence to identify the exact modulation scheme in use.

\end{itemize} 

{\bf Outline.} The rest of this paper is organized as follows. Section-II introduces the system model and provides the necessary background on THz channel model. Section-III provides the framework for the proposed modulation mode detection method. Section-IV describes the proposed feature-based modulation classification method in detail. Section-V provides extensive numerical results followed by discussions. Finally, Section-VI concludes the paper.

\section{System Model \& THz Channel Model}
\label{sec:sys-model}

\subsection{System Model} 
\begin{figure}[ht]
\begin{center}
	\includegraphics[width=3.3in]{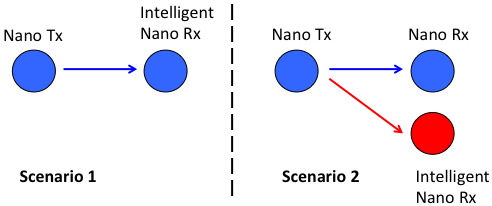} 
\caption{The system model: two distinct application scenarios for the considered intelligent nano receiver.}
\label{fig:sys-model}
\end{center}
\end{figure}

We consider an intelligent/cognitive nano receiver listening to a signal of interest in THz band. With the objective of modulation mode detection and modulation classification, two distinct application scenarios are foreseeable: 1) the nano receiver is the intended recipient of the signal transmitted by the nano transmitter; 2) the nano receiver is not the intended recipient; rather, it overhears the signal intended for some other nano receiver in the nearby vicinity (see Fig. \ref{fig:sys-model}). Note that the application scenario 1 (scenario 2) implies that the nano receiver has a trustworthy (untrustworthy) relationship with the nano transmitter. Scenario 2 depicts passive interception which arises in military, defense, and security applications.

\subsection{The THz Channel Model}
The THz channel is extremely frequency-selective due to absorption by the water molecules. Water vapors constitute the main factor altering the Terahertz channel. As the THz signal propagates through the channel, it is attenuated due to the resonance of some molecules in the atmosphere at specific frequencies. Specifically, the THz channel impulse response $h(t)$ is given as \cite{Josep:TWC:2011},\cite{Josep:TC:2014}:
\begin{equation}
\label{eq:THzh(t)}
h(t) = h_{ant}^T(t)\ast h_c(t) \ast h_{ant}^R(t)
\end{equation}
where $h_c(t)=\mathcal{F}^{-1}\{H_c(f)\}$ ($\mathcal{F}^{-1}\{.\}$ is the inverse Fourier transform operator) where the channel frequency response $H_c(f)$ (a.k.a the frequency-dependent pathloss) is:
\begin{equation}
H_c(f) = H_{spread}(f)H_{abs}(f),
\end{equation}
where the spreading loss $H_{spread}(f)$ (due to expansion of a wave as it travels through the medium) is:
\begin{equation}
H_{spread}(f)=\frac{1}{\sqrt{4\pi d^2}}\exp{\bigg(-j\frac{2\pi fd}{c}\bigg)},
\end{equation}
where $d$ is the distance between the nano transmitter and nano receiver, $c=3.8\times 10^8$ m/sec is the speed of light in free-space. The molecular absorption loss $H_{abs}(f)$ is:
\begin{equation}
H_{abs}(f)=\exp{(-\frac{k(f)d}{2})},
\end{equation}
where $k(f)$ is the medium absorption coefficient given as:
\begin{equation}
k(f) = \frac{p}{p_0}\frac{\mathbb{T}_{stp}}{\mathbb{T}} \sum_i Q^i\sigma^i(f)
\end{equation}
where $p$ is the system pressure, $p_0$ is the reference pressure (i.e., 1 atm), $\mathbb{T}$ is the system temperature, $\mathbb{T}_{stp}$ is the temperature at standard pressure (i.e., $\mathbb{T}_{stp}=273.15$ K), $Q^i$ is the number of molecules per volume unit of gas $i$ and $\sigma^i(f)$ is the absorption cross-section of gas $i$.  Using the radiative transfer theory \cite{Goody:Book:1995}, and given a set of parameters, absorption lines $k(f)$ can be modeled at any temperature and pressure from HITRAN database \cite{HITRAN:2017}. 

The antenna impulse response in transmission is:
\begin{equation}
\label{eq:hantT}
h_{ant}^T(t) = \frac{\partial}{\partial t} \int \mathbf{J}(t,(x,y)) dV,
\end{equation}
where $V$ stands for the volume occupied by the antenna, $\mathbf{J}$ (units of [1/$m^2$sec]) is the current distribution on the antenna due to an impulse input current $\delta(t)$ in the
transverse direction relative to the observation direction. The current distribution on the antenna 
$J$ depends on the particular antenna design, and usually it can only be numerically obtained (e.g., in COMSOL \cite{comsol}). Nevertheless, in plain words, Eq. (\ref{eq:hantT}) implies that the radiated electromagnetic field is proportional to the first time derivative of the current density at the antenna surface. 

Finally, the antenna impulse response in reception is proportional to the time integral of the antenna impulse response in transmission:
\begin{equation}
h_{ant}^R(t) = \int_0^{t} h_{ant}^T(\tau) d\tau
\end{equation}

\section{Modulation Mode Detection}
\label{sec:PL}

\subsection{Motivation}
A nano-scale communication system operating in THz band could either utilize pulse based modulation (PBM), or, carrier based modulation (CBM) \cite{Ian:ICC:2012},\cite{Josep:TC:2014}. Under the classic CBM approach, the nano transmitter and nano receiver both tune to a center frequency where absorption loss due to atmospheric molecules is minimum; then, standard modulation schemes---phase shift keying (PSK), quadrature amplitude modulation (QAM)---are used. The PBM approach, on the other hand, relies upon transmission and subsequent successful reception of extremely short-lived (few hundreds of femto-seconds long) pulses to realize simple modulation schemes (e.g., on-off keying, amplitude shift keying etc.). 

We consider a situation where the nano receiver does not know apriori the mode of modulation (PBM, or, CBM) used by the nano transmitter. Such situation could arise, for example, when the pair of nano devices constitutes a hybrid communication system whereby the nano transmitter is capable of switching between PBM and CBM on slot basis. This in turn necessitates that the nano receiver should build a systematic mechanism to detect the mode of modulation upon reception of the signal during every slot. We dub this problem as modulation mode detection problem, and solve it by constructing a binary hypothesis test at the nano receiver. 

\subsection{The Binary Hypothesis Test}
We assume a time-slotted system with $T$ seconds long time-slots. We further assume that the nano transmitter utilizes a Gaussian pulse $p(t)$ for pulse based modulation \cite{Ian:ICC:2012},\cite{Josep:TC:2014}, and a raised-cosine (RC) pulse $q(t)$ for carrier based modulation. Specifically, $p(t)=\frac{1}{\sqrt{2\pi \sigma_p^2}}\exp{\bigg(-\frac{(t-\mu_p)^2}{2\sigma_p^2}\bigg)}=a\exp{\bigg(-\frac{(t-b)^2}{2c^2}\bigg)}$ where $a=1/(\sqrt{2\pi \sigma_p^2})$, $b=\mu_p$, $c=\sigma_p$; $a,b,c>0$ specify the amplitude, the center of the pulse, and the spread of the pulse respectively. Inline with the previous work \cite{Ian:ICC:2012},\cite{Josep:TC:2014}, we assume that the Gaussian pulse $p(t)$ is of duration $c=T_p<T$. The RC pulse is given as: $q(t)=\text{sinc} (t/T)\frac{\cos \pi \alpha (t/T)}{1-(2\alpha t/T)^2}$ where $0\leq \alpha <1$ is the roll-off factor which models the excess bandwidth, and $T$ is the symbol duration (thus, one slot conveys only one symbol). 

Let $u_0(t)=\sum_k p(t-kT)$, $u_1(t)=(\sum_k b[k]q(t-kT))\cos (2\pi f_c t+\phi)$ be the signal transmitted from the nano transmitter under PBM approach, and CBM approach respectively. $b[k]$ is the (PSK, QAM) symbol transmitted during the $k$-th time-slot, $f_c$ is the carrier frequency, and $\phi$ is the phase of the carrier wave, under the CBM approach. Let $r(t)$ be the passband signal received at the nano receiver\footnote{The proposed binary hypothesis test for modulation mode detection, though performed on received passband signals, is equally valid for a baseband setup. The only notable (yet minor) distinctions are that the test statistic will have $2N$ degrees of freedom and the baseband PBM signal will have zero imaginary part, but everything else remains intact.}. Then, the modulation mode detection problem is formulated as the following binary hypothesis test during the observation interval of length $\beta$ time-slots, i.e., $T_o=\beta T$:
\begin{equation}
	\label{eq:H0H1}
	 \begin{cases} H_0: & \text{(PBM)} \; r(t)=s_0(t)+w(t) \\
                                 H_1: & \text{(CBM)} \; r(t)=s_1(t)+w(t) \end{cases}
\end{equation}
where $s_0(t)=u_0(t)\ast h(t)=\sum_k f(t-kT)$, where $f(t)=p(t)\ast h(t)$. And, $s_1(t)=u_1(t)\ast h(t)=(\sum_k b[k]g(t-kT))\cos (2\pi f_c t+\phi)$, where $g(t)=q(t)\ast h(t)$. The $h(t)$ is the impulse response of the THz channel given by Eq. (\ref{eq:THzh(t)}). Finally, $w(t)$ is the additive white Gaussian noise (AWGN) process with auto-correlation function $E[w(t_1)w(t_2)]=\sigma^2\delta(t_1-t_2)$ where $\delta(t_1-t_2)$ is the Kronecker-delta function, and $\sigma^2$ is the power spectral density (PSD) of AWGN process. 

We collect $N$ samples of $r(t)$ during each slot; therefore, we have the following equivalent test in discrete-time:
\begin{equation}
	\label{eq:H0H1}
	 \begin{cases} H_0: & \text{(PBM)} \; r[n]=s_0[n]+w[n] \\
                                 H_1: & \text{(CBM)} \; r[n]=s_1[n]+w[n] \end{cases}
\end{equation}
where $r[n]=r(nT_s)$ with $T/T_s=N$, and $n=0,1,...,N-1$. Also, $w[n]\sim\mathcal{N}(0,\sigma^2)$ for $n=0,1,...,N-1$ are independent and identically distributed (i.i.d.). Then, conditioned on $s_i[n]$, we have: $r[n]|H_i\sim \mathcal{N}(s_i[n],\sigma^2)$ where $i \in \{0,1\}$. Let $\tilde{r}[n]=r[n]/\sigma$. Then, $\tilde{r}[n]|H_i\sim \mathcal{N}(s_i[n],1)$. 

The assumption $T_p<T$ (i.e., low duty cycle of $u_0(t)$) is the key to observe that the energy in $u_0(t)$ is less than the energy in $u_1(t)$ during any observation interval of length $T_o$. Therefore, the modulation mode detection problem boils down to an energy detection problem. Let $A=\sum_{n=0}^{N-1} (\tilde{r}[n])^2$. Then, $A$ follows a non-central chi-squared distribution with $N$ degrees of freedom. Furthermore, $A|H_0$ ($A|H_1$) has non-centrality parameter $\lambda_0=\sum_{n=0}^{N-1}(s_0[n])^2$ ($\lambda_1=\sum_{n=0}^{N-1}(s_1[n])^2$). Thus, a simple test statistic could be constructed as follows: 
\begin{equation}
\label{eq:teststat}
\mathcal{T} = \frac{\sigma^2}{N} A = \frac{\sigma^2}{N} \sum_{n=0}^{N-1} (\tilde{r}[n])^2 \underset{H_0}{\overset{H_1}{\gtrless}} \eta
\end{equation}
where $\eta$ is the comparison threshold, a design parameter. 

\subsection{Performance}
The hypothesis test in (\ref{eq:teststat}) incurs two kinds of errors, the type-I errors (wrongly declaring CBM), and the type-II errors (wrongly declaring PBM). The probability of type-I error is: 
\begin{equation}
\begin{split}
P_{e,1} &= P(H_1|H_0) = P(\mathcal{T} > \eta|H_0) = P(A > N\eta/\sigma^2|H_0) \\
&= 1-P(A \leq N\eta/\sigma^2|H_0) = 1 - F_A(N\eta/\sigma^2,N,\lambda_0)
\end{split}
\end{equation}
where $F_X(x,m,\lambda)=P(X\leq x)$ is the cumulative distribution function (CDF) of a non-central chi-squared distributed random variable (R.V.) $X$ with $m$ degrees of freedom, and non-centrality parameter $\lambda$. Specifically,  $F_X(x,m,\lambda)=\exp{(-\lambda/2)}\sum_{j=0}^{\infty} \frac{(\lambda/2)^j}{j!} F_Y(x,m+2j)$ where $F_Y(x,l)$ is the CDF of a central chi-squared distributed R.V. $Y$ with $l$ degrees of freedom. Furthermore, $F_Y(x,l)=\Phi(x/2,l/2)$ where $\Phi(x,l)$ is the regularized lower gamma function defined as $\Phi(x,l)=\frac{\gamma (x,l)}{\Gamma (l)}$, where $\gamma (x,l)$ is the lower incomplete gamma function and $\Gamma (l)$ is the gamma function. Alternatively, the CDF of $X$ could also be written as: $F_X(x,m,\lambda)=1-Q_{m/2}(\sqrt{\lambda},\sqrt{x})$; $Q_{\zeta}(\xi,\nu)$ is the generalized Marcum-Q function (with $\xi,\nu \in \mathbb{R}^+$ and $\zeta \in \mathbb{I}^+$). Therefore, the probability of type-I error is given as:
\begin{equation}
P_{e,1}=Q_{N/2}(\sqrt{\lambda_0},\sqrt{\eta^{'}})=Q_{N/2} \bigg( \sqrt{\sum_{n=0}^{N-1}(s_0[n])^2},\sqrt{\frac{N\eta}{\sigma^2}} \bigg)
\end{equation}
where $\eta^{'}=\frac{N\eta}{\sigma^2}$. 

The probability of type-II error is given as: 
\begin{equation}
\begin{split}
P_{e,2} &= P(H_0|H_1) = P(\mathcal{T} < \eta|H_1) = P(A < N\eta/\sigma^2|H_1) \\
&= F_A(N\eta/\sigma^2,N,\lambda_1)=1-Q_{N/2}(\sqrt{\lambda_1},\sqrt{\eta^{'}}) \\
&=1-Q_{N/2} \bigg( \sqrt{\sum_{n=0}^{N-1}(s_1[n])^2},\sqrt{\frac{N\eta}{\sigma^2}} \bigg) 
\end{split}
\end{equation}

\section{Modulation Classification for Carrier based Modulation Schemes}
\label{sec:EH}

When modulation mode detection declares that the nano transmitter of interest has employed carrier based modulation during current time-slot, then a subsequent task of an intelligent nano receiver is to decide which particular modulation scheme (among $M$-PSK, and $M$-QAM) has been used. 

In this section, we discuss the specifics of the proposed modulation classification scheme. Specifically, we first represent the received signal as a Gaussian mixture model (GMM), and utilize Expectation Maximization (EM) algorithm to learn the GMM parameters from the training data in a systematic manner. We then do Gaussian approximation of each GMM, and utilize the symmetric Kullback-Leibler divergence to systematically identify the modulation scheme in use.

\subsection{Gaussian Mixture Model Representation of the Received Signal}
Under Gaussian mixture model, the probability density function (pdf) $u(x)$ of the (observed) mixture random variable $U$ defined over a probability space $x\in \mathcal{X}$ is the convex/weighted sum of the $Q$ component pdfs:
\begin{equation}
\label{eq:gmm}
u(x) = \sum_{q=1}^Q \pi_q \phi_q(x)
\end{equation}
where each $\phi_q(x)$ is a Gaussian pdf which satisfies: $\phi_q(x)\geq 0$, $\int_{x\in \mathbb{R}^d} \phi_q(x)dx=1$. The weights/priors satisfy: $\pi_q\geq 0$, $\sum_{q=1}^Q \pi_q=1$. $d$ is the dimension of the data $x$; $d=1$ for BPSK, and $d=2$ for $M$-PSK and $M$-QAM schemes (with $M>2$). 

Next, a quick example to illustrate the representation of the received signal via GMM. For a received signal containing BPSK modulation with 0's and 1's being equally likely and 0 mapped to -1$V$ and 1 mapped to +1$V$, the GMM will have: $Q=2$, $\pi_0=\pi_1=0.5$, $\phi_0(x)\sim N(\mu_0,\Sigma_0)$, $\phi_1(x)\sim N(\mu_1,\Sigma_1)$ with $\mu_0=-1$ and $\mu_1=+1$.

\subsection{Learning the GMM parameters via the Expectation-Maximization Algorithm}

The GMM has $3Q$ unknown parameters which are learned using iterative Expectation-Maximization algorithm applied on training data $\{x_m\}_{m=1}^{M}$.
The posterior probability for each point $x_m$ in the training data (i.e., the likelihood of $x_m$ belonging to component $q$ of the mixture) is computed as follows ($j$ is the iteration number):
\begin{equation}
\label{eq:em1}
p_{m,q}^{(j)} = \frac{ \pi_q^{(j)} \phi_q(x_m, \mu_q^{(j)}, \Sigma_q^{(j)}) } {\sum_{\hat{q}=1}^{Q} \pi_{\hat{q}}^{(j)} \phi(x_m, \mu_{\hat{q}}^{(j)}, \Sigma_{\hat{q}}^{(j)})}
\end{equation}
The $Q$ number of priors are updated as follows:
\begin{equation}
\pi_q^{(j+1)} = \frac{1}{M} \sum_{m=1}^{M} p_{m,q}^{(j)}
\end{equation}
The $Q$ number of means are updated as follows:
\begin{equation}
\mu_q^{(j+1)} = \frac{ \sum_{m=1}^{M} p_{m,q}^{(j)} x_m } { \sum_{m=1}^{M} p_{m,q}^{(j)} }
\end{equation}
The $Q$ number of (co-)variances are updated as follows:
\begin{equation}
\label{eq:em4}
\Sigma_q^{(j+1)} = \frac{ \sum_{m=1}^{M} p_{m,q}^{(j)} (x_m-\mu_q^{(j)})(x_m-\mu_q^{(j)})^T } { \sum_{m=1}^{M} p_{m,q}^{(j)} }
\end{equation}
The iterative EM algorithm monotonically increases the objective (likelihood) function value, and is said to converge when the increase in likelihood function value between two successive iterations becomes less than a very small threshold $\epsilon$.

\subsection{Least-Squares based Estimation of the THz Channel Impulse Response \& Compensation via Deconvolution}

The aforementioned representation of the received signal as a GMM, and subsequent learning of the GMM parameters via the EM algorithm is valid for a noise-limited channel only. For communication in THz band, one needs to explicitly estimate the channel impulse response (CIR) (of Eq. (\ref{eq:THzh(t)})), and then compensate it before applying the GMM+EM based framework. To this end, Appendix A presents a Least-Squares (LS) based method for THz channel impulse response estimation. Let $\hat{h}(t)$ denote the obtained LS channel estimate, then the deconvolution procedure to remove the effect of $h(t)$ from the received signal $r(t)$ works as follows. Let $R(f)=\mathcal{F}(r(t))$, $\hat{H(f)}=\mathcal{F}(\hat{h}(t))$ ($\mathcal{F}(.)$ denotes the Fourier transform operator). Then, the effect of THz CIR is compensated for by: $\hat{r}(t)=\mathcal{F}^{-1}(\frac{R(f)}{\hat{H}(f)})$. The proposed GMM+EM based framework for modulation classification is then applied on ($M$) samples of $\hat{r}(t)$.

\subsection{Symmetric Kullback-Leibler Divergence as the Sole Feature for Modulation Classification}

Kullback-Leibler divergence (KLD) is a (directional) measure of the distance between the two probability density functions $y(x)$ and $z(x)$ sharing a common probability space $\mathcal{X}$:
\begin{equation}
D(y||z)  = \int_{\mathcal{X}} y(x) \log\frac{y(x)}{z(x)}dx
\end{equation}
Note that $D(y||y)=0$. 

Let us now represent $y$ and $z$ as the two Gaussian mixture densities:
\begin{equation}
   \begin{aligned}
 y(x) = \sum_{q=1}^Q \pi_{y,q} \phi(x, \mu_{y,q}, \Sigma_{y,q})   \\
 z(x) = \sum_{q^{'}=1}^{Q^{'}} \pi_{z,q^{'}} \phi(x, \mu_{z,q^{'}}, \Sigma_{z,q^{'}})
  \end{aligned}
\end{equation}
where $y$ represents the signal received from the nano transmitter of interest, and $z$ represents the template signal in the database. 

Directly computing the KLD between two mixture densities $y$ and $z$ is very involved; therefore, we follow an alternate approach where we replace $y$ and $z$ with their Gaussian approximations $\hat{y}$ and $\hat{z}$ respectively \cite{Hershey:ICASSP:2007}. Under this approach, the mean and (co-)variance of $\hat{y}$ are given as \cite{Hershey:ICASSP:2007}: 
 \begin{equation}
 \label{eq:Gaussmu}
  \mu_{\hat{y}} = \sum_{q=1}^{Q} \pi_{y,q} \mu_{y,q}
  \end{equation}
\begin{equation}
\label{eq:Gausssigma}
\Sigma_{\hat{y}} = \sum_{q=1}^{Q}  \pi_{y,q} \left[ \Sigma_{y,q} + 
(\mu_{y,q} - \mu_{\hat{y}})(\mu_{y,q} - \mu_{\hat{y}})^T \right] 
\end{equation}
The expressions for mean $\mu_{\hat{z}}$ and variance $\Sigma_{\hat{z}}$ of $\hat{z}$ could be written in a similar manner.

With this, the KLD between the two (approximately) Gaussian (mixture) densities is given as:

\begin{equation}
 \begin{aligned}
&D(\hat{y}||\hat{z})=\\                      
&\frac{1}{2}   \left[ \log\left( \frac{|\Sigma_{\hat{z}}|}{|\Sigma_{\hat{y}}|}\right)  + \text{Tr}  \left[ \Sigma_{\hat{z}}^{-1} \Sigma_{\hat{y}} \right] +(\mu_{\hat{y}}-\mu_{\hat{z}})^T\Sigma_{\hat{z}}^{-1}(\mu_{\hat{y}}-\mu_{\hat{z}}) - d \right] \\
\label{8} 
  \end{aligned}
\end{equation}
where $\text{Tr}(.)$ and $|.|$ represent the trace, and determinant of a matrix respectively. One can again verify that $D(\hat{y}||\hat{y})=0$.

Due to the fact that $D(\hat{y}||\hat{z})\neq D(\hat{z}||\hat{y})$, we utilize the symmetric KLD instead for decision-making for modulation classification. The symmetric KLD is defined as: 
\begin{equation}
\label{eq:Dsym}
D_{sym}(\hat{y}||\hat{z}) = 0.5 ( D(\hat{y}||\hat{z}) + D(\hat{z}||\hat{y}) )
\end{equation}

\subsection{The Proposed Method for Modulation Classification}

We construct an offline database at the nano receiver which contains the constellation points of the following modulation schemes $\mathcal{M}=\{\text{BPSK,QPSK,8-PSK,16-QAM}\}$. Next, we construct template signals $\tau_i(t)$, $i=1,...,|\mathcal{M}|$, and assume a noise-limited channel to represent each of them as a GMM ($|.|$ represents the cardinality of a set). The template signal $\tau_i(t)$ then serves as the ground truth under hypothesis $i$ which states that the received signal of interest utilizes the modulation scheme on $i$-th index of $\mathcal{M}$. Ultimately, we compute the $|\mathcal{M}|$ symmetric KLDs, between the Gaussian approximation of the GMM representing the received deconvolved signal $\hat{r}(t)$ and the Gaussian approximation of the GMM representing $\tau_i(t)$, $i=1,...,|\mathcal{M}|$. The index $i$ for which the symmetric KLD is the minimum is utilized to pick the corresponding modulation scheme from the set $\mathcal{M}$ to declare it as the modulation scheme used by the nano transmitter of interest during the current time-slot.

The proposed algorithm for modulation classification as implemented by the intelligent nano receiver is summarized below:

\begin{enumerate}

\item Obtain the least-squares estimate $\hat{h}(t)$ of the THz CIR by plugging the (samples of) received signal $r(t)$ and the known training symbols in Eq. (\ref{eq:lsest}) (in Appendix A).

\item Obtain the deconvolved (noise-limited only) signal as: $\hat{r}(t)=\mathcal{F}^{-1}(\frac{R(f)}{\hat{H}(f)})$.

\item Represent $\hat{r}(t)$ as a Gaussian mixture model as in Eq. (\ref{eq:gmm}), and learn the GMM parameters via the iterative EM algorithm using Eqs. (\ref{eq:em1})-(\ref{eq:em4}).

\item Approximate the GMM pdf due to $\hat{r}(t)$ as well as the GMM pdfs due to the template signals $\tau_i(t)$ in the database each as a Gaussian pdf using Eqs. (\ref{eq:Gaussmu}),(\ref{eq:Gausssigma}).

\item Compute the $|\mathcal{M}|$ symmetric KLDs (using Eq. (\ref{eq:Dsym})), between the Gaussian approximation of the GMM representing the received deconvolved signal $\hat{r}(t)$ and the Gaussian approximation of the GMM representing $\tau_i(t)$, $i=1,...,|\mathcal{M}|$. The index $i$ for which the symmetric KLD is the minimum is utilized to pick the corresponding modulation scheme from the set $\mathcal{M}$ to declare it as the modulation scheme in use.

\end{enumerate}

\section{Numerical Results}
\label{sec:results}

\subsection{The THz channel Setup}
Let $H(f)=\mathcal{F}\{h(t)\}$ represent the channel frequency response; therefore, $H(f)=H_{ant}^T(f) H_c(f) H_{ant}^R(f)$. Inline with previous work \cite{Josep:TC:2014}, we assume that $|H_{ant}^T(f) H_{ant}^R(f)|=\frac{\lambda_0}{\sqrt{4\pi}}$ which implies that point-dipole antennas are deployed by both nano transmitter and nano receiver. $\lambda_0=c/f_0$ where $f_0$ is the antenna design frequency---basically the center frequency of the pulse power spectral density---$f_0 = 1.6$ THz. 

\begin{figure}[ht]
\begin{center}
	\includegraphics[width=3.6in]{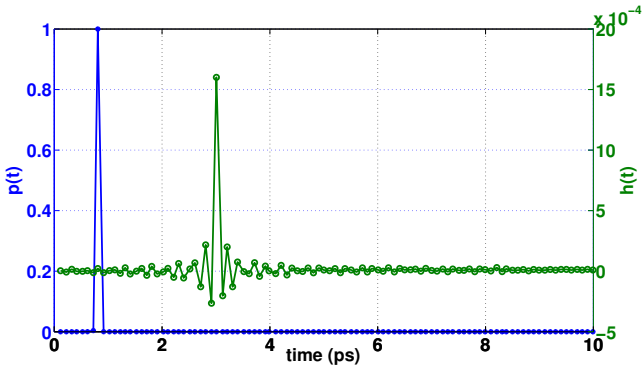} 
\caption{ THz channel impulse response $h(t)$ observed at the nano receiver when a 100 femto seconds long Gaussian pulse $p(t)$ is transmitted by the nano transmitter at $t=800$ fs. Additionally, to obtain this plot, the nano receiver was placed at a distance of 1 mm from the nano transmitter \cite{Josep:TC:2014}.}
\label{fig:cir}
\end{center}
\end{figure}

Under the above setting, Fig. \ref{fig:cir} plots the THz channel impulse response $h(t)$ observed by the nano receiver when a 100 femto seconds long Gaussian pulse $p(t)$ is transmitted by the nano transmitter at $t=800$ fs. For this plot, we placed the nano receiver at a distance of 1 mm from the nano transmitter \cite{Josep:TC:2014}. The channel impulse response shows that the transmitted pulse is delayed and spread in time (which could potentially lead to inter-symbol interference). We utilize the CIR of Fig. \ref{fig:cir} to simulate the performance of modulation mode detection and modulation classification in the sequel.

\subsection{Simulation Results: Modulation Mode Detection}

{\it Remark 1:} We note that the essence of modulation mode detection is energy detection; i.e., the transmitted CBM signal has more energy than the transmitted PBM signal during any observation interval. And, this fact holds even when the transmitted signal passes through the THz channel. Therefore, for all the results on mode detection below, we do hypothesis testing on the raw received signal in the passband (in other words, we do not explicitly compensate for the THz CIR).

\begin{figure}[ht]
\begin{center}
	\includegraphics[width=3.8in]{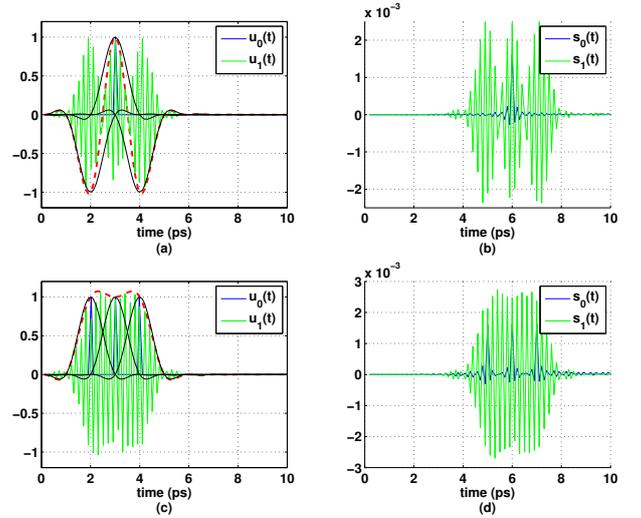} 
\caption{Simulation setup for Mode detection: PBM approach utilizes on-off keying, while CBM approach utilizes BPSK scheme. For CBM approach, a carrier of frequency $5$ THz was used. In Figs. \ref{fig:sim_setup}(a)-(b), the transmitted symbol sequence is $\{-1,+1,-1\}$, while in Figs. \ref{fig:sim_setup}(c)-(d), the transmitted symbol sequence is $\{+1,+1,+1\}$. Furthermore, in Figs. \ref{fig:sim_setup}(a), (c), the red curve represents the envelope $e_{u_1}=\sum_k b[k]q(t-kT)$ of $u_1(t)$, while each of the black curve represents $q_k(t)=q(t-kT)$ for $k \in \{2,3,4\}$. Finally, the duty cycle of the Gaussian pulses is set to 10\% under PBM approach. }
\label{fig:sim_setup}
\end{center}
\end{figure}

Fig. \ref{fig:sim_setup} graphically summarizes the essential waveforms (and hence the simulation setup) for the mode detection. To obtain Fig. \ref{fig:sim_setup}, we set $\alpha=0.8$, $T=1$ ps, $T_o=3T$, $\sigma_p=20$ fs, $a=1$, and $N=40$. Furthermore, we used a carrier wave with center frequency $f_c=5$ THz. For PBM, on-off keying (where nothing is sent for logic 0) was used, while for CBM, BPSK scheme (thus $\phi=0$) was used. The transmitted symbols are $\{-1,+1,-1\}$.  

\begin{figure}[ht]
\begin{center}
	\includegraphics[width=3.8in]{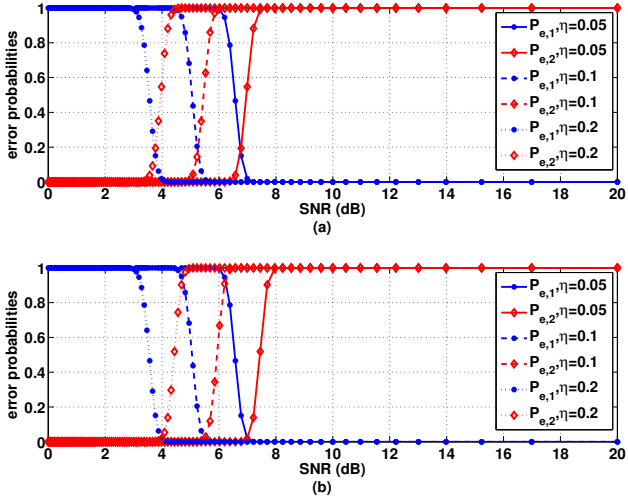} 
\caption{Mode detection: probabilities of error vs. SNR. For Fig. \ref{fig:PevsSNR} (a), the transmitted symbol sequence is $\{-1,+1,-1\}$, while for Fig. \ref{fig:PevsSNR} (b), the transmitted symbol sequence is $\{+1,+1,+1\}$.}
\label{fig:PevsSNR}
\end{center}
\end{figure}

Fig. \ref{fig:PevsSNR} (a), (b) plots the two error probabilities $P_{e,1}$ (wrongly declaring CBM), $P_{e,2}$ (wrongly declaring PBM) against the signal-to-noise ratio (SNR) for three different values of the threshold $\eta$, for the symbol sequence $\{-1,+1,-1\}$, $\{+1,+1,+1\}$ respectively\footnote{We define SNR as SNR$=1/\sigma^2$; the SNR here is merely an indicator of quality of measurements.}. Note that $P_{e,2}$ is the CDF of the R.V. $A=\sum_{n=0}^{N-1} (\tilde{r}[n])^2$, while $P_{e,1}$ is the complementary CDF (CCDF) of $A$; therefore, $P_{e,1}$ ($P_{e,2}$) decays (grows) with increase in $\eta^{'}=\frac{N\eta}{\sigma^2}$. This implies that, with $N$ and $\eta$ fixed, one cannot minimize both errors for all values of SNR. This is verified by Fig. \ref{fig:PevsSNR} as for a fixed value of $\eta$, there is only one (pareto-) optimal SNR $SNR_{opt}$ where both errors are jointly minimized, i.e., $P_{e,1}=P_{e,2}=P_{e,min}$. Specifically, in Fig. \ref{fig:PevsSNR} (a), for $\eta=0.05$, $\eta=0.1$, $\eta=0.2$, we have $SNR_{opt}=6.8$ dB, $SNR_{opt}=5.5$ dB, $SNR_{opt}=3.9$ dB respectively, and $P_{e,min}=0.2$. While, in Fig. \ref{fig:PevsSNR} (b), for $\eta=0.05$, $\eta=0.1$, $\eta=0.2$, we have $SNR_{opt}=7.1$ dB, $SNR_{opt}=5.7$ dB, $SNR_{opt}=4$ dB respectively, and $P_{e,min}=0.01$. Thus, $P_{e,min}$ depends upon specific symbol sequence $\{b[k]\}$ actually sent from the nano transmitter.

\begin{figure}[ht]
\begin{center}
	\includegraphics[width=3.8in]{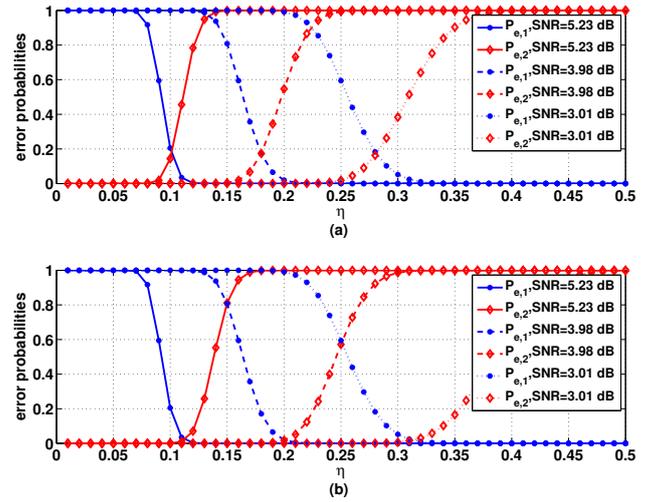} 
\caption{Mode detection: probabilities of error vs. $\eta$. For Fig. \ref{fig:PevsEta} (a), the transmitted symbol sequence is $\{-1,+1,-1\}$, while for Fig. \ref{fig:PevsEta} (b), the transmitted symbol sequence is $\{+1,+1,+1\}$.}
\label{fig:PevsEta}
\end{center}
\end{figure}

Fig. \ref{fig:PevsEta} (a), (b) plots the two error probabilities against $\eta$ for three different values of the SNR, for the symbol sequence $\{-1,+1,-1\}$, $\{+1,+1,+1\}$ respectively. Once again, for a fixed value of SNR, there is only one (pareto-) optimal threshold $\eta_{opt}$ where both errors are jointly minimized, i.e., $P_{e,1}=P_{e,2}=P_{e,min}$. Specifically, in Fig. \ref{fig:PevsEta} (a), for SNR$=5.23$ dB, SNR$=3.98$ dB, SNR$=3.01$ dB, we have $\eta_{opt}=0.1$, $\eta_{opt}=0.18$, $\eta_{opt}=0.28$ respectively, and $P_{e,min}=0.2$. While, in Fig. \ref{fig:PevsEta} (b), for SNR$=5.23$ dB, SNR$=3.98$ dB, SNR$=3.01$ dB, we have $\eta_{opt}=0.11$, $\eta_{opt}=0.21$, $\eta_{opt}=0.32$ respectively, and $P_{e,min}=0.01$. Once again, $P_{e,min}$ depends upon specific symbol sequence $\{b[k]\}$ actually sent from the nano transmitter.

\subsection{Simulation Results: Modulation Classification}

{\it Remark 2:} We note that the proposed GMM+EM based modulation classification framework applies to noise-limited signals only. Therefore, for all the results on modulation classification below, we first obtain the deconvolved signal $\hat{r}(t)$ by compensating for the THz CIR using the LS based method presented in Appendix A.

Fig. \ref{fig:const} shows the constellation plots for each of the modulation schemes considered in this work. We utilized Fig. \ref{fig:const} to construct the database $\mathcal{M}$ which contains the constellation points for each of the four modulation schemes considered. The constellation points of $i$-th modulation scheme, in turn, become the $Q$ means of the GMM representation of the template signal $\tau_i(t)$.

\begin{figure}[ht]
\begin{center}
	\includegraphics[width=3.6in]{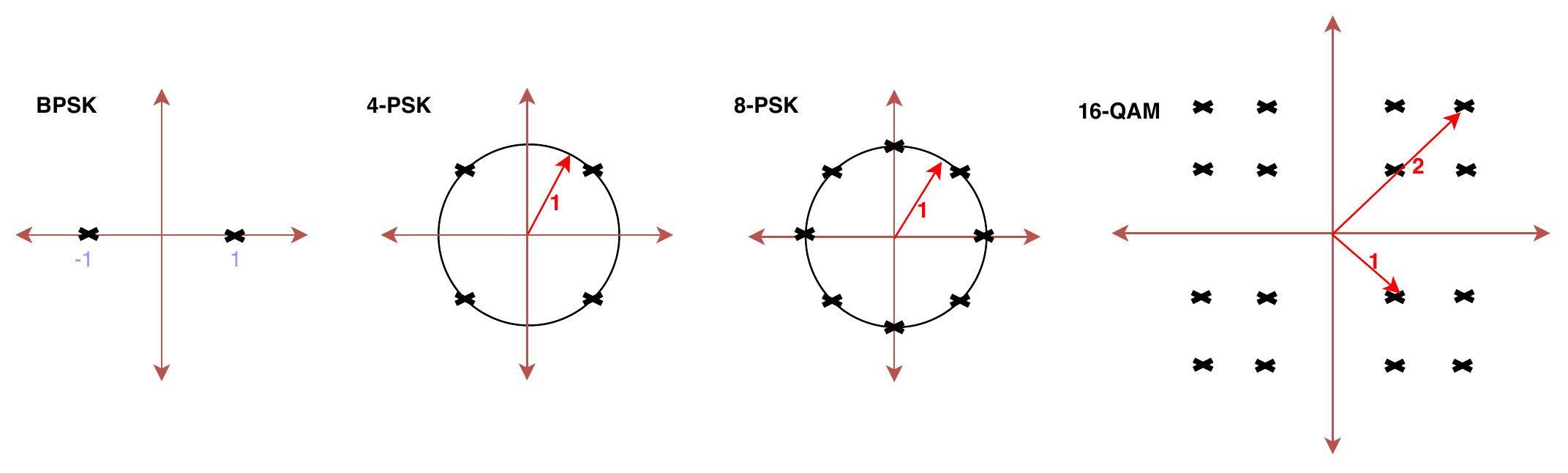} 
\caption{ Labelled constellation plots for the modulation schemes considered in this work.}
\label{fig:const}
\end{center}
\end{figure} 

\begin{figure}[ht]
\begin{center}
	\includegraphics[width=3.8in]{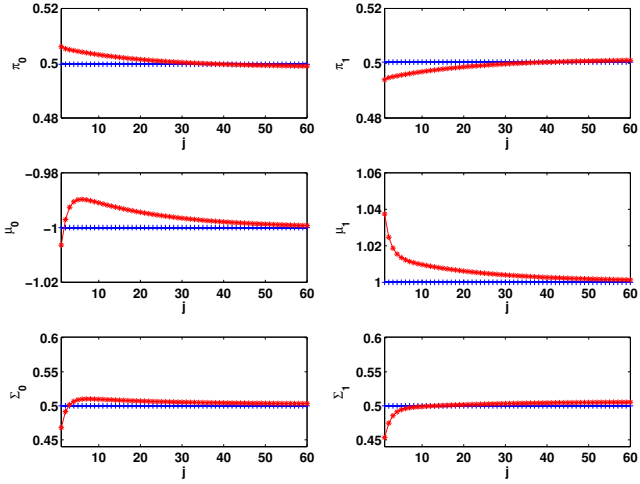} 
\caption{EM algorithm learns the true parameters of the GMM representing BPSK scheme ($Q=2$) in about 60 iterations. Blue curve represents the true value of the parameters, while the red curves represent the iterates of the EM algorithm. For this result, we used training data of size $M=1e5$ samples, and convergence threshold $\epsilon=1e-4$.}
\label{fig:gmmem}
\end{center}
\end{figure}

Fig. \ref{fig:gmmem} illustrates the performance of the EM algorithm in the noise-limited channel (i.e., the EM algorithm is trained over the deconvolved signal $\hat{r}(t)$ to learn the GMM parameters). The figure corresponds to the case where the GMM represents the BPSK modulation scheme. Specifically, the true parameters for the training data were: $\beta_0=0.5$, $\beta_1=0.5$, $\mu_0=-1$, $\mu_1=1$, $\Sigma_0=0.5$, $\Sigma_0=0.5$, while we initialized the EM algorithm with the following guess at iteration 0: $\hat{\beta}_0(0)=0.6$, $\hat{\beta}_1(0)=0.4$, $\hat{\mu}_0(0)=-1.2$, $\hat{\mu}_1(0)=1.3$, $\hat{\Sigma}_0(0)=0.4$, $\hat{\Sigma}_1(0)=0.6$. We observe that the EM algorithm learns the $3Q$ number of parameters of the GMM model for BPSK scheme ($d=1$) very efficiently (in about 60 iterations).

\begin{figure}[ht]
\begin{center}
	\includegraphics[width=3.8in]{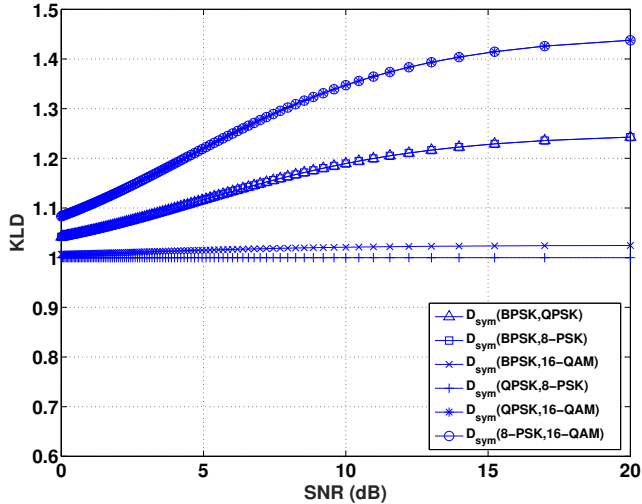} 
\caption{Pair-wise symmetric KLDs for the considered modulation schemes against SNR. Note that the curve $D_{sym}$(QPSK,16-QAM) is superimposed on $D_{sym}$(8-PSK,16-QAM), and the curve $D_{sym}$(BPSK,QPSK) is superimposed on $D_{sym}$(BPSK,8-PSK).}
\label{fig:kld_snr}
\end{center}
\end{figure}

Fig. \ref{fig:kld_snr} plots all the pair-wise symmetric KLDs for the considered modulation schemes against the SNR. For this figure, an ideal situation is considered where the GMM parameters for all the modulation schemes are perfectly known. Fig. \ref{fig:kld_snr} illustrates that all the pair-wise symmetric KLDs show a monotonic (non-decreasing) trend with increase in SNR which is an intuitively pleasing result. This is because all the self-KLDs are zero regardless of the SNR value\footnote{Given two pdfs $y$ and $z$, we call $D(y||y)$ the self-KLD, while we call $D(y||z)$  the (directional) pair-wise KLD. By invoking Eq. (\ref{eq:Dsym}), one can have pair-wise symmetric KLD.}; therefore, the distance between a self-KLD and the corresponding pair-wise symmetric KLDs increases with increase in SNR. The overall takeaway message by Fig. \ref{fig:kld_snr} is that the symmetric KLD is indeed a viable (and sufficient) feature for modulation classification. 

\begin{figure}[ht]
\begin{center}
	\includegraphics[width=3.8in]{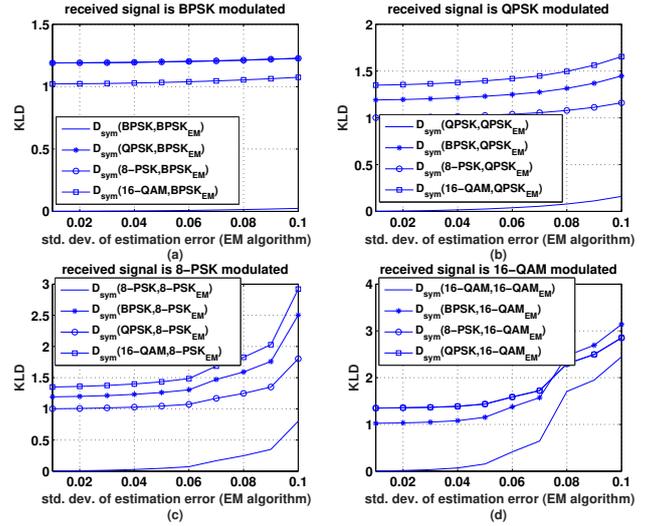} 
\caption{Self-KLD and corresponding pair-wise symmetric KLDs for each of the four considered modulation schemes plotted against the (standard deviation of) estimation error of the EM algorithm. For this plot, number of Monte-Carlo simulations done was $1e4$. Furthermore, we set $\sigma=0.1$. For illustration, $D_{sym}(\text{QPSK,BPSK}_{EM})$ implies the pair-wise symmetric KLD between the Gaussian approximated pdf of the GMM of the template signal in the database representing QPSK modulation, and the Gaussian approximated pdf of the GMM of the received deconvolved signal containing BPSK modulation. Furthermore, the GMM parameters of the template signal are perfectly known, while the GMM parameters of the received deconvolved signal are learned via the EM algorithm.}
\label{fig:kld_em_err}
\end{center}
\end{figure}

Fig. \ref{fig:kld_em_err} plots the self-KLD and corresponding pair-wise symmetric KLDs for each of the four considered modulation schemes against the (standard deviation of) estimation error of the EM algorithm. Specifically, given a received deconvolved signal $\hat{r}(t)$ (whose GMM parameters are learned by the EM algorithm), the KLDs between $\hat{r}(t)$ and all the template signals $\tau_i(t)$, $i=1,...,|\mathcal{M}|$ in the database (whose GMM parameters are perfectly known to the nano receiver) are computed. Under this setting, Fig. \ref{fig:kld_em_err} considers the training data $\{x_m\}_{m=1}^M$ with increasing noise levels, and thus, investigates the impact of noisy estimation of the GMM parameters (of the received signal of interest) by the EM algorithm on the self-KLD and all the pair-wise symmetric KLDs. We observe that the values of both the self-KLDs as well as the pair-wise symmetric KLDs increase with increase in estimation error\footnote{Let $y$ be the Gaussian approximated pdf of the GMM with perfectly known parameters (representing one of the template signals from the database, say, BPSK), and let $\hat{y}$ be the Gaussian approximated pdf of the GMM (representing the received deconvolved signal containing BPSK modulation) whose parameters are estimated by the EM algorithm, then the self-KLD $D(y||\hat{y})\neq 0$.}. But most importantly, the gap between any self-KLD and the corresponding pair-wise symmetric KLDs remains constant. This in turn implies that the proposed method could correctly classify the modulation scheme in use very efficiently, even in the presence of large estimation errors by the EM algorithm. 

\begin{figure}[ht]
\begin{center}
	\includegraphics[width=3.8in]{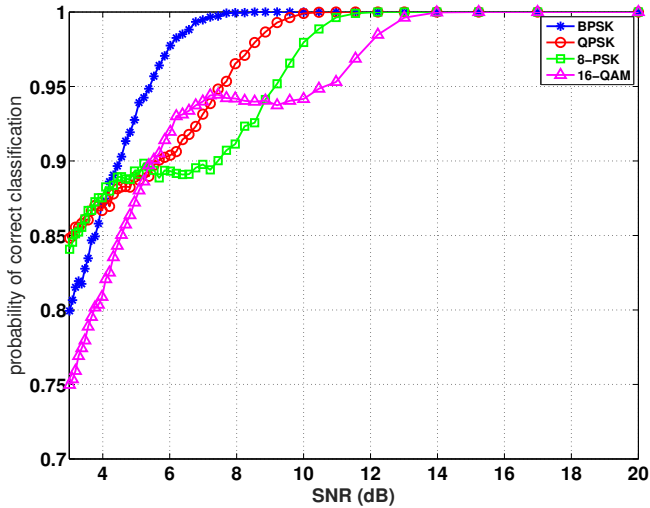} 
\caption{$P_{cc}$ vs. SNR. For this plot, number of Monte-Carlo simulations done was $2e4$.}
\label{fig:pcc}
\end{center}
\end{figure}

Finally, Fig. \ref{fig:pcc} plots the main result---probability of correct classification $P_{cc}$ (correctly declaring $i$-th modulation scheme in $\mathcal{M}$) against the SNR. As expected, $P_{cc}$ for all modulation schemes increases with increase in SNR, and converges to the maximum value of 1 for moderate SNR values (about $8-14$ dB). However, we note that, for a pre-specified $P_{cc}$, the required SNR for higher-order constellations schemes is larger and vice versa (this is again the anticipated result because higher-order constellations are successfully decoded at higher SNRs only). The reason for such behavior is that for the higher order modulation schemes (8-PSK, and 16-QAM), the gap between the self-KLD and the corresponding pair-wise symmetric KLDs reduces slightly with increase in estimation error of the EM algorithm (see Fig. \ref{fig:kld_em_err} (c),(d)) which in turn reduces the $P_{cc}$ for higher order modulation schemes at low SNRs.

\section{Conclusion}
\label{sec:conclusion}

We considered the problem of designing an intelligent/cognitive nano receiver operating in THz band. Specifically, we investigated two essential ingredients of an intelligent nano receiver---modulation mode detection, and modulation classification. For modulation mode detection, we constructed a binary hypothesis test in nano-receiver's passband, and provide closed-form expressions for the two error probabilities. As for modulation classification, we first did explicit least-squares based THz channel estimation and subsequent compensation via deconvolution. We then represented the received deconvolved signal by a Gaussian mixture model, and learned its parameters via Expectation-Maximization algorithm. We then did Gaussian approximation of each mixture density to compute symmetric Kullback-Leibler divergence in order to differentiate between various modulation schemes (i.e., $M$-ary phase shift keying, $M$-ary quadrature amplitude modulation). Extensive simulation results were performed which attested to the effectiveness of the two proposed methods. 

This work opens up many possibilities for the follow-up work on intelligent nano-receivers, e.g., modulation prediction as well as modulation classification using machine learning tools; design of matched filters at the nano receiver for combating the inter-symbol interference and for modulation mode detection; investigation of cyclo-stationarity based approaches for mode detection etc.

\appendices

\section{Least-Squares based THz Channel Impulse Response Estimation}
\label{app:cir_est}

The symbols $b[k]$ sent by the nano transmitter and the signal $r[k]$ received by the nano receiver are related as: $r[k]=b[k]\ast h[k]+w[k]=\sum_{l=0}^{L} h[l]b[k-l]+w[k]$. This work considers a training-based, least-squares approach to THz CIR estimation. Let $\{b[k]\}_{k=k_1}^{k=k_m}$ denote the training symbols. Define: $\mathbf{h}=[h[0],...,h[L]]^T$; basically $\mathbf{h}$ contains the $L+1$ taps of the CIR. Furthermore, let $\mathbf{r}=[r[k_1+L],r[k_1+L+1],...,r[k_m]]^T$, $\mathbf{w}=[w[k_1+L],w[k_1+L+1],...,w[k_m]]^T$. Then: 
\begin{equation}
{\mathbf{B}} = 
 \begin{pmatrix}
  b[k_1+L] & b[k_1+L-1] & \cdots & b[k_1] \\
  b[k_1+L+1] & b[k_1+L] & \cdots & b[k_1+1] \\
  \vdots  & \vdots  & \ddots & \vdots  \\
  b[k_m] & b[k_m-1] & \cdots & b[k_m-L] 
 \end{pmatrix}
\end{equation}

Then, the received signal at the nano receiver can be compactly written in matrix-vector form as:
\begin{equation}
{\mathbf{r}} = {\mathbf{B}}{\mathbf{h}} + {\mathbf{w}}
\end{equation}

Finally, the least-squares (LS) based estimate of the THz CIR is given as:
\begin{equation}
\label{eq:lsest}
\hat{\mathbf{h}} = ({\mathbf{B}^H}{\mathbf{B}})^{-1}{\mathbf{B}^H}{\mathbf{r}} 
\end{equation}

Specifically, $\hat{\mathbf{h}} \sim \mathcal{CN}({\mathbf{h}},{\mathbf{\Sigma}_h})$ where ${\mathbf{\Sigma}_h} = \sigma^2({\mathbf{B}^H}{\mathbf{B}})^{-1}$. 

It is worth mentioning that the proposed least-squares based solution exists only when $\mathbf{B}$ is a full column-rank matrix. In other words, the length of training data should be: $k_m-k_1 \geq 2L$.

\footnotesize{
\bibliographystyle{IEEEtran}
\bibliography{references}

\begin{thebibliography}{10}
\providecommand{\url}[1]{#1}
\csname url@rmstyle\endcsname
\providecommand{\newblock}{\relax}
\providecommand{\bibinfo}[2]{#2}
\providecommand\BIBentrySTDinterwordspacing{\spaceskip=0pt\relax}
\providecommand\BIBentryALTinterwordstretchfactor{4}
\providecommand\BIBentryALTinterwordspacing{\spaceskip=\fontdimen2\font plus
\BIBentryALTinterwordstretchfactor\fontdimen3\font minus
  \fontdimen4\font\relax}
\providecommand\BIBforeignlanguage[2]{{%
\expandafter\ifx\csname l@#1\endcsname\relax
\typeout{** WARNING: IEEEtran.bst: No hyphenation pattern has been}%
\typeout{** loaded for the language `#1'. Using the pattern for}%
\typeout{** the default language instead.}%
\else
\language=\csname l@#1\endcsname
\fi
#2}}

\bibitem{Akyildiz:Elsevier:2010}
I.~F. Akyildiz and J.~M. Jornet, ``Electromagnetic wireless nanosensor
  networks,'' \emph{Nano Communication Networks (Elsevier) Journal}, vol.~1,
  no.~1, pp. 3--19, Mar. 2010.

\bibitem{Ian:Nanonetworks:2008}
I.~F. Akyildiz, F.~Brunetti, and C.~Bl{\'a}zquez, ``Nanonetworks: A new
  communication paradigm,'' \emph{Computer Networks}, vol.~52, no.~12, pp.
  2260--2279, 2008.

\bibitem{Mona:BodyNets:2015}
\BIBentryALTinterwordspacing
M.~Nafari and J.~M. Jornet, ``Metallic plasmonic nano-antenna for wireless
  optical communication in intra-body nanonetworks,'' in \emph{Proceedings of
  the 10th EAI International Conference on Body Area Networks}, ser. BodyNets
  '15.\hskip 1em plus 0.5em minus 0.4em\relax ICST, Brussels, Belgium, Belgium:
  ICST (Institute for Computer Sciences, Social-Informatics and
  Telecommunications Engineering), 2015, pp. 287--293. [Online]. Available:
  \url{http://dx.doi.org/10.4108/eai.28-9-2015.2261410}
\BIBentrySTDinterwordspacing

\bibitem{Josep:TWC:2011}
J.~M. Jornet and I.~F. Akyildiz, ``Channel modeling and capacity analysis for
  electromagnetic wireless nanonetworks in the terahertz band,'' \emph{IEEE
  Transactions on Wireless Communications}, vol.~10, no.~10, pp. 3211--3221,
  October 2011.

\bibitem{Qammer:TTHz:2016}
Q.~H. Abbasi, H.~E. Sallabi, N.~Chopra, K.~Yang, K.~Qaraqe, and A.~Alomainy,
  ``Terahertz channel characterisation inside the human skin at the
  nano-scale,'' \emph{IEEE Transactions on THz Science and Technology}, vol.~6,
  no.~3, pp. 427 -- 434, May, 2016.

\bibitem{Gupta:Globecom:2015}
A.~Gupta, M.~Medley, and J.~M. Jornet, ``Joint synchronization and symbol
  detection design for pulse-based communications in the thz band,'' in
  \emph{Global Communications Conference (GLOBECOM), 2015 IEEE}.\hskip 1em plus
  0.5em minus 0.4em\relax IEEE, 2015, pp. 1--7.

\bibitem{Ian:ICC:2012}
R.~G. Cid-Fuentes, J.~M. Jornet, I.~F. Akyildiz, and E.~Alarc{\'o}n, ``A
  receiver architecture for pulse-based electromagnetic nanonetworks in the
  terahertz band,'' in \emph{Communications (ICC), 2012 IEEE International
  Conference on}.\hskip 1em plus 0.5em minus 0.4em\relax IEEE, 2012, pp.
  4937--4942.

\bibitem{Josep:TC:2014}
J.~M. Jornet and I.~F. Akyildiz, ``Femtosecond-long pulse-based modulation for
  terahertz band communication in nanonetworks,'' \emph{IEEE Transactions on
  Communications}, vol.~62, no.~5, pp. 1742--1754, May 2014.

\bibitem{Jornet:Elsevier:2012}
J.~M. Jornet, J.~Capdevila-Pujol, and J.~Sole-Pareta, ``Phlame: A physical
  layer aware mac protocol for electromagnetic nanonetworks in the terahertz
  band,'' \emph{Nano Communication Networks (Elsevier) Journal}, vol.~3, no.~1,
  pp. 74 -- 81, 2012.

\bibitem{Han:CN:2013}
C.~Han, J.~M. Jornet, E.~Fadel, and I.~F. Akyildiz, ``A cross-layer
  communication module for the internet of things,'' \emph{Computer Networks},
  vol.~57, no.~3, pp. 622--633, 2013.

\bibitem{Iwaszczuk:PhDThesis:2012}
K.~Iwaszczuk, ``Terahertz technology for defense and security-related
  applications,'' \emph{PhD thesis, Technical University of Denmark}, vol. 172,
  2012.

\bibitem{MOSHIR:NCN:2016}
\BIBentryALTinterwordspacing
F.~Moshir and S.~Singh, ``Modulation and rate adaptation algorithms for
  terahertz channels,'' \emph{Nano Communication Networks}, vol.~10, pp. 38 --
  50, 2016, terahertz Communications. [Online]. Available:
  \url{http://www.sciencedirect.com/science/article/pii/S1878778916300412}
\BIBentrySTDinterwordspacing

\bibitem{Octavia:IETComm:2007}
O.~A. Dobre, A.~Abdi, Y.~Bar-Ness, and W.~Su, ``Survey of automatic modulation
  classification techniques: classical approaches and new trends,'' \emph{IET
  Communications}, vol.~1, no.~2, pp. 137--156, April 2007.

\bibitem{Wang:Globecom:2012}
J.~G. Liu, X.~Wang, J.~Nadeau, and H.~Lin, ``Modulation classification based on
  gaussian mixture models under multipath fading channel,'' in \emph{2012 IEEE
  Global Communications Conference (GLOBECOM)}, Dec 2012, pp. 3970--3974.

\bibitem{Wu:TWC:2008}
H.~C. Wu, M.~Saquib, and Z.~Yun, ``Novel automatic modulation classification
  using cumulant features for communications via multipath channels,''
  \emph{IEEE Transactions on Wireless Communications}, vol.~7, no.~8, pp.
  3098--3105, August 2008.

\bibitem{Peng:WOCC:2017}
S.~Peng, H.~Jiang, H.~Wang, H.~Alwageed, and Y.~D. Yao, ``Modulation
  classification using convolutional neural network based deep learning
  model,'' in \emph{2017 26th Wireless and Optical Communication Conference
  (WOCC)}, April 2017, pp. 1--5.

\bibitem{Ali:PhyComm:2017}
\BIBentryALTinterwordspacing
A.~Ali and F.~Yangyu, ``Unsupervised feature learning and automatic modulation
  classification using deep learning model,'' \emph{Physical Communication},
  vol.~25, pp. 75 -- 84, 2017. [Online]. Available:
  \url{http://www.sciencedirect.com/science/article/pii/S1874490717300435}
\BIBentrySTDinterwordspacing

\bibitem{Zhu:WileyCh6:2014}
\BIBentryALTinterwordspacing
Z.~Zhu and A.~K. Nandi, \emph{Machine Learning for Modulation
  Classification}.\hskip 1em plus 0.5em minus 0.4em\relax Wiley-Blackwell,
  2014, ch.~6, pp. 81--95. [Online]. Available:
  \url{https://onlinelibrary.wiley.com/doi/abs/10.1002/9781118906507.ch6}
\BIBentrySTDinterwordspacing

\bibitem{Vavouris:MOCAST:2018}
A.~K. Vavouris, F.~D. Dervisi, V.~K. Papanikolaou, and G.~K. Karagiannidis,
  ``An energy efficient modulation scheme for body-centric nano-communications
  in the thz band,'' in \emph{2018 7th International Conference on Modern
  Circuits and Systems Technologies (MOCAST)}, May 2018, pp. 1--4.

\bibitem{Zhang:Access:2017}
R.~Zhang, K.~Yang, Q.~H. Abbasi, K.~A. Qaraqe, and A.~Alomainy, ``Analytical
  characterisation of the terahertz in-vivo nano-network in the presence of
  interference based on ts-ook communication scheme,'' \emph{IEEE Access},
  vol.~5, pp. 10\,172--10\,181, 2017.

\bibitem{Goody:Book:1995}
R.~M. Goody and Y.~L. Yung, \emph{Atmospheric radiation: theoretical
  basis}.\hskip 1em plus 0.5em minus 0.4em\relax Oxford University Press, 1995.

\bibitem{HITRAN:2017}
\BIBentryALTinterwordspacing
I.~Gordon \emph{et~al.}, ``The hitran 2016 molecular spectroscopic database,''
  \emph{Journal of Quantitative Spectroscopy and Radiative Transfer}, vol. 203,
  pp. 3 -- 69, 2017, hITRAN2016 Special Issue. [Online]. Available:
  \url{http://www.sciencedirect.com/science/article/pii/S0022407317301073}
\BIBentrySTDinterwordspacing

\bibitem{comsol}
\BIBentryALTinterwordspacing
``Comsol multiphysics software - understand, predict, and optimize.'' [Online].
  Available: \url{https://www.comsol.com/comsol-multiphysics}
\BIBentrySTDinterwordspacing

\bibitem{Hershey:ICASSP:2007}
J.~R. Hershey and P.~A. Olsen, ``Approximating the kullback leibler divergence
  between gaussian mixture models,'' in \emph{2007 IEEE International
  Conference on Acoustics, Speech and Signal Processing - ICASSP '07}, vol.~4,
  April 2007, pp. IV--317--IV--320.

\end{thebibliography}
}

\vfill\break

\end{document}